\documentclass[aps,pra,preprint,groupedaddress]{revtex4-2}
\usepackage{graphicx,caption,subcaption,amsmath}
\renewcommand{\varphi}{\wp} % define \varphi to be \wp (Weierstrass rho)
\begin{document}

% Use the \preprint command to place your local institutional report
% number in the upper righthand corner of the title page in preprint mode.
% Multiple \preprint commands are allowed.
% Use the 'preprintnumbers' class option to override journal defaults
% to display numbers if necessary
%\preprint{}

%Title of paper
\title{On alleged solutions of the cubically nonlinear Schr\"odinger equation}

% repeat the \author .. \affiliation  etc. as needed
% \email, \thanks, \homepage, \altaffiliation all apply to the current
% author. Explanatory text should go in the []'s, actual e-mail
% address or url should go in the {}'s for \email and \homepage.
% Please use the appropriate macro foreach each type of information

% \affiliation command applies to all authors since the last
% \affiliation command. The \affiliation command should follow the
% other information
% \affiliation can be followed by \email, \homepage, \thanks as well.
\author{Hans Werner Sch\"urmann}
\email[]{hwschuer@uos.de}
%\altaffiliation{}
\affiliation{Department of Mathematics, Computer Science, Physics\\ University of Osnabr\"uck, Germany}

\author{Valery Serov}
\email[]{vserov@cc.oulu.fi,valserov@gmail.com}
%\altaffiliation{}
\affiliation{Research Unit  of Mathematical Sciences\\ University of Oulu, Finland}
%Moscow Centre of Fundamental and Applied Mathematics-\\ - Lomonosov Moscow State University, Russia}

%Collaboration name if desired (requires use of superscriptaddress
%option in \documentclass). \noaffiliation is required (may also be
%used with the \author command).
%\collaboration can be followed by \email, \homepage, \thanks as well.
%\collaboration{}
%\noaffiliation

%\date{}
\begin{abstract}
On the basis of analytical results, we present a numerical example that indicates inconsistency of a widely used ansatz with cubically nonlinear Schr\"odinger equation.
\end{abstract}

% insert suggested PACS numbers in braces on next line
% insert suggested keywords - APS authors don't need to do this
%\keywords{}

%\maketitle must follow title, authors, abstract, \pacs, and \keywords
\maketitle

% body of paper here - Use proper section commands
% References should be done using the \cite, \ref, and \label commands
%\section{}
% Put \label in argument of \section for cross-referencing
%\section{\label{}}
%\subsection{}
%\subsubsection{}

\section{Introduction}

As is well-known, the cubically nonlinear Schr\"odinger equation (CNLSE) (see Eq.(1a) below)
%$$
% i A_t (x,t)+pA_{xx}(x,t)+qA(x,t)|A(x,t)|^2=0
%$$
is an appropriate model to study a great variety of physical settings in, e.g., nonlinear optics \cite{AkhA}, hydrodynamics \cite{Z}, and plasma physics \cite{H}. Due to this variety, the actual meaning of the independent variables $x,t$ and of parameters $p$ and $q$ is different for different problems.

As an integrable differential equation, the CNLSE possesses an infinite set of conservation laws and can be solved by general techniques (e.g., IST, Darboux transformation).

In general, it is not easy to derive practical solutions for certain physically relevant problems using these general techniques. Thus, special ansatz (trial) solutions have been suggested. In the present article we consider one of them, originally presented in \cite{AEK} and reconsidered in \cite{C}. With reference to \cite{AEK}, in a recent article, we have claimed inconsistency of this ansatz \cite{Se}. Since a modification of \cite{AEK} has been proposed in \cite{C}, the cause of the present article is to check whether this modification is leading to a correction of our claim in \cite{Se}. -- We think that it is important to know whether this wide used ansatz \cite{Footnote} is suitable as a solution of the CNLSE.

After stating the problem in Section II, we transform, in Section III, the basic system $\{(6a), (6b)\}$ (see below) to a dynamical system, that is suitable for the final consistency check in Section V. In Section IV we present solutions for $d(t)$ and $Q(x,t)$ (see Eq.(1b) below), necessary for use in Section V. We conclude with some comments  in Section VI.

\section{Statement of the problem}

We consider the basic system considered in \cite{C}
\begin{subequations}
\begin{align}
 i A_t (x,t)+pA_{xx}(x,t)+qA(x,t)|A(x,t)|^2=0,\\
 A(x,t)=(Q(x,t)+id(t)e^{i\phi(t)},
\end{align}
\end{subequations}
%$$
%(a)\quad i A_t (x,t)+pA_{xx}(x,t)+qA(x,t)|A(x,t)|^2=0, 
%$$
%\begin{equation}
%(b)\quad A(x,t)=(Q(x,t)+id(t)e^{i\phi(t)}, 
%\end{equation}
where $x,t,p,q,Q,d,\phi$ are real and $d_t(t)\ne 0,Q_x(x,t)\ne 0$ (generic case). The consistency of the basic system (1) represents the problem.

Usually, the CNLSE (1a) is written in the form \cite{AEK}
\begin{equation}
i A_t (x,t)+A_{xx}(x,t)+2A(x,t)|A(x,t)|^2=0.
\end{equation}
Obviously, with $p=1,q=2$, Eq.(2) is a particular case of CNLSE (1a), hence results obtained w.r.t. \cite{C} also refer to \cite{AEK}.

In the next Section, assuming validity of the basic system, we derive relations leading to a transformed system in terms of solutions $d=\sqrt{z},Q$ that must hold necessarily for the basic system. -- Finally, the validity assumption leads to a contradiction.

\section{Transformation of the basic system to a dynamical system}

If the basic system is assumed to be consistent, the ansatz (1b) must satisfy the CNLSE (1a). Hence, if $A(x,t)$ is substituted into (1a), we obtain (following the line indicated in \cite{AEK}, \cite{C}, \cite{Se}), separating imaginary and real parts (assuming $p=1$)

\begin{subequations}
\begin{align}
Q_t(x,t)=d(t)\left(\phi_t(t)-q(d^2(t)+Q^2(x,t))\right)     \label{eq:diffusionpartialdiff_boundaries0}\\
(Q_x(x,t))^2=-\frac{q}{2}Q^4(x,t)-(q^2d^2(t)-\phi_t(t))Q^2(x,t)-2d_t(t)Q(x,t)+b(t),\label{eq:diffusionpartialdiff_boundariesRs}
\end{align}
\end{subequations}
%$$
%(a)\quad Q_t(x,t)=d(t)\left(\phi_t(t)-q(d^2(t)+Q^2(x,t))\right)     
%$$     
%\begin{equation}          
%(b)\quad (Q_x(x,t))^2=-\frac{q}{2}Q^4(x,t)-(q^2d^2(t)-\phi_t(t))Q^2(x,t)-2d_t(t)Q(x,t)+b(t),
%\end{equation}
where $b(t)$ denotes the first $x-$integral of the real part of CNLSE from (1). In order to eliminate (unknown) $b(t)$ and $\phi_t(t), d_t(t)$ from this system it is useful to transform it (see \cite{AEK}, \cite{C}) as follows.

Evaluation of the integrability condition $Q_{xt}=Q_{tx}$ leads to three ordinary differential equations (see Eq.(6) in \cite{C} and Eqs.(7a)-(7c) in \cite{Se}) that can be integrated successively, hence
\begin{subequations}
\begin{align}
\phi_t(t)=-2qd^2(t)+c_1,\\
b(t)=\frac{1}{4}(2c_2-2c_1d^2(t)+3qd^4(t)),\\
(d_t(t))^2=-4q^2d^6(t)+4qc_1d^4(t)-(c_1^2+2qc_2)d^2(t)+c_3, 
\end{align}
\end{subequations}
%$$
%(a)\quad \phi_t(t)=-2qd^2(t)+c_1
%$$
%$$
%(b)\quad b(t)=\frac{1}{4}(2c_2-2c_1d^2(t)+3qd^4(t))
%$$
%\begin{equation}
%(c) \quad (d_t(t))^2=-4q^2d^6(t)+4qc_1d^4(t)-(c_1^2+2qc_2)d^2(t)+c_3,
%\end{equation}
where $c_1,c_2$ and $c_3$ are integration constants. Setting $d^2(t)=z(t)$, Eq.(4c) can be rewritten as
\begin{equation}
(z_t(t))^2=\alpha_1z^4(t)+4\beta_1z^3(t)+6\gamma_1z^2(t)+4\delta_1z(t)+\epsilon_1=:R_1(z),
\end{equation}
where
$$
\alpha_1=-16q^2, \quad \beta_1=4qc_1,\quad \gamma_1=-\frac{2}{3}(c_1^2+4qc_2),\quad \delta_1=c_3,\quad \epsilon_1=0.
$$
By using Eqs.(4a) and (4b), the dynamical system \{(3a), (3b)\} reads 
\begin{subequations}
\begin{align}
Q_t(x,t)=\sqrt{z(t)}(c_1-q(3z(t)+Q^2(x,t))),\\
(Q_x(x,t))^2=-\frac{q}{2}Q^4(x,t)+(c_1-3qz(t))Q^2(x,t)+\frac{z_t(t)}{\sqrt{z(t)}}Q(x,t)+2c_2+\frac{3}{2}qz^2(t)-c_1z(t)
:= R_2(x,t).
\end{align}
\end{subequations}
%$$
%(a)\quad Q_t(x,t)=\sqrt{z(t)}(c_1-q(3z(t)-Q^2(x,t)))
%$$
%\begin{equation}
%(b) \quad Q_x(x,t))^2=-\frac{q}{2}Q^4(x,t)+(c_1-3qz(t))Q^2(x,t)+\frac{z_t(t)}{\sqrt{z(t)}}Q(x,t)+2c_2+\frac{3}{2}qz^2(t)-c_1z(t)
%\end{equation}
With solution $z(t)$ of Eq.(5) this system is valid necessarily if the basic system is assumed to be consistent. We note that Eqs.(5), (6a), (6b) are equivalent with the corresponding equations in \cite{AEK} and \cite{C}, namely, with Eqs.(13), (5), (15) (due to (4a) in \cite{Se} and $c_1=W,c_2=\frac{H}{2}, c_3=D$) in \cite{AEK}, and with Eqs.(7), (4a), (5) in \cite{C} ($c_1=-\omega_0, c_2=\frac{k_2}{2}, c_3=k_1, p=1,q=2$).

%The next Section is devoted to possible solutions $z(t)$ and $Q(x,t)$. 

\section{solutions of Eq.(5) and Eq.(6$b$)}

Obviously Eqs.(5) and (6b) are of the same type. Solutions can be obtained by applying Weierstrass' solution $y(x)$ (Eq.(11) in \cite{Se}). In general, $y(x)$ might be complex and unbounded. 
%For the numerical test in the next Section only real and bounded solutions are relevant. 
The solution of Eq.(5) reads (the prime denotes differentiation w.r.t. $z$)
$$
z(t)=z_0+
$$
\begin{equation}
+\frac{\frac{1}{2}R'_1(z_0)\left(\wp(t;g_{2z},g_{3z})-\frac{1}{24}R''_1(z_0)\right)\pm\frac{d\wp(t;g_{2z},g_{3z})}{dt}\sqrt{R_1(z_0)}+\frac{1}{24}R_1(z_0)R_1'''(z_0)}{2\left(\wp(t;g_{2z},g_{3z})-\frac{1}{24}R_1''(z_0)\right)^2-\frac{1}{48}R_1(z_0)R_1''''(z_0)},
\end{equation} 
where $z_0=z(0)$ and 
$$
g_{2z}=\frac{4}{3}(c_1^2+4qc_2)^2-16qc_1c_3,
$$
$$
g_{3z}=\frac{8}{27}\left(54q^2c_3^2-18qc_1c_3(c_1^2+4qc_2)+(c_1^2+4qc_2)^3\right).
$$
Equation (6b) is solved by (here the prime denotes differentiation w.r.t. $Q$)
$$
Q(x,t)=Q_0(t)+
$$
\begin{equation}
+\frac{\frac{1}{2}R'_2(Q_0)\left(\wp(x;g_{2Q},g_{3Q})-\frac{1}{24}R_2''(Q_0)\right)\pm\frac{d \wp(x;g_{2Q},g_{3Q})}{dx}\sqrt{R_2(Q_0)}+\frac{1}{24}R_2(Q_0)R_2'''(Q_0)}{2\left(\wp(x;g_{2Q},g_{3Q})-\frac{1}{24}R_2''(Q_0)\right)^2-\frac{1}{48}R_2(Q_0)R_2''''(Q_0)},
\end{equation}
with
$$
g_{2Q}=\frac{c_1^2}{12}-qc_2,\quad g_{3Q}=-\frac{(c_1-3qz(t))}{216}\left(c_1^2-24qc_1z(t)+36q^2z^2(t)+36qc_2\right)+\frac{qz^2_t(t)}{32z(t)}.
$$
%We firstly note that (5) and (6b) are necessary if the basic system is assumed to be consistent. Secondly we emphasize that (7) and (8), and onlyb (7) and (8), are the solutions of Eqs.(5) and (6b), respectively (see  \cite{W}). 
%Thirdly we note that $Q_0:=Q(0,t)=Q_0(t)$ depends on $t$ in general, since (8) has been derived by integration w.r.t. $x$.

%By  the above mentioned (see end of Section III) equivalence of Eq.(5) with Eq.(13) in \cite{AEK}, Eq.(7) is equivalent with Eq.(22) in \cite{AEK}. Correspondingly, Eq.(8) is equivalent with Eq.(24) in \cite{AEK}.

\section{Inconsistency of the basic system}

By the foregoing transformation, system $\{(6a),(6b)\}$ represents vanishing imaginary and real parts of Eq.(1a), respectively. With (necessarily) valid (7) (as the only solution of Eq.(5)) and (8) (as the only solution of Eq.(6b)) system $\{(6a),(6b)\}$ must be considered. Solution (7) is valid for any real (see Eqs.(4a)-(4c)) parameters $c_j, q$ and any real (we consider real $g_{2Q}$, $g_{3Q}$) $z_0$. Solution (8) is valid for any parameters $c_j, q$ and any $z_0, Q_0$; as the unique solution of (6b) in system $\{(6a),(6b)\}$, (8) is the only solution that must satisfy (6a), if the system is assumed to be consistent:
\begin{equation}
P(x, t;c_j,q,z_0,Q_0):=Q_t(x,t)-\sqrt{z(t)}(c_1-q(3z(t)+Q^2(x, t)))=0.
\end{equation}
Since (9) must hold for any real $x,t,c_j,q,z_0$ and any $Q_0$, we select at random $x=1,t=1,q=-1,c_1=-2,c_2=0.4, c_3=0.13,z_0=1,Q_0=1$ leading to  
$$
P(1,1;c_j,q,1,1)=0.113,
$$
contrary to (9): Since (9) is necessary for consistency of the basic system, this system is inconsistent.

\section{Concluding Comments}

Whereas ansatz (1b) is compatible with the CNLSE (1a) in the nongeneric cases (as outlined in \cite{Se}, Section IV), above we have presented arguments that this is not the case in the generic case.
 -- In view of the results in \cite{AEK} and \cite{C}, we might be mistaken in this claim. However, if we assume that it is correct, the question arises as to why the inconsistency was not uncovered in \cite{AEK} and \cite{C} (may be, the cause for \cite{C} was a certain discomfort with \cite{AEK}) or in publications based on \cite{AEK} (we did not find citations of \cite{C}). In this respect, we point out again that systems $\{(5), (6)\}$ in \cite{AEK}, $\{(4a), (5)\}$ in \cite{C}, and system $\{(6a), (6b)\}$ are (only) necessary for validity of the basic system. 
As far as we can see, neither in \cite{AEK} nor in \cite{C} it has been checked whether the real part of Eq.(1a) is vanishing if the 
corresponding solution $\{z(t), Q(x,t)\}$ is substituted (in \cite{AEK}: $z(t)$ and $Q(x,t)$ according to (22) and (24), respectively, into Eq.(5); in \cite{C}: $z(t)$ and $Q(x,t)$, according to (9) and (13)-(15), respectively, into Eq.(4a)). -- This might be a reason, why the error has not been detected.

A further question arises as to whether a solution $\tilde{Q}(x,t)\ne Q(x,t)$ is possible. As mentioned in Section I, a formally different (different from $Q$ in (4) and different from $Q$ according to (8)) solution $\tilde{Q}(x,t)$ has been suggested in \cite{C}.
 Instead of integrating Eq.(5) in \cite{C} (Eq.(6) in \cite{AEK}), the author of \cite{C} preferred to transform Eq.(4a) in \cite{C} to a Riccati equation (13) leading to $A(x,t)$ according to Eq.(16) in \cite{C}. If $\tilde{Q}(x,t)$, according to (13)-(15) in \cite{C} is assumed to be the correct solution of Eq.(4a) in \cite{C} (equivalent to Eq.(6a)), it must satisfy Eq.(5), which is equivalent 
to Eq.(6b) due to the transformation of Section 3. However, Eq.(6b) has the unique solution (8) by construction (see \cite{W}). Hence, $\tilde{Q}(x,t)=Q(x,t)$. -- As mentioned above, Eq.(5) in \cite{C} has not been verified with solution $\tilde{Q}(x,t)$ according to (13)-(15) (as far as we can see).

Finally, we would like to make some comments on the genesis of the present article.

Originally, the foregoing contents was part of a Comment on \cite{CMKTA} submitted to Phys. Rev. A (in 2022). Unsurprisingly, the authors of \cite{CMKTA} strongly recommended rejection of the comment with a "substantiated" reply [arXiv.22.09.05892], containing reference \cite{VSN} as a confirmation of their claim (Ref.\cite{C} in the Reply). After several rounds of reports, replies, and appeals the Editor of the Phys. Rev. A finally rejected our Comment. Due to some recommendations in the review process we submitted an upblowed version of the Comment as a regular article to the Journal of Mathematical Physics. In short, the review process again led to rejection with almost the similar arguments of missing novelty. Admittedly, in a strict sense, the claim of nonexistence of certain solutions of an important differential equation is not a claim of new solutions. Nevertheless, it is, in our opinion, an interesting question whether nonexistent solutions of the NLSE are suitable to model physical phenomena with the NLSE using these solutions. 
It seems that the authors of \cite{CMKTA} and \cite{VSN} have ignored this aspect. They simply relied on the solutions presented in \cite{AEK} (Eqs.(22), (24)), but did not check Eq.(5) in \cite{AEK}, equivalent with Eq.(9) above that is not satisfied in the generic case.

In view of the wide-spread use of the alleged solutions presented above, we could not leave the two rejections without reaction: The result is presented in \cite{Se} and the present article. -- The solutions used in \cite{CMKTA} (even if they claimed in "good agreement between theory and experiment"\cite{VSN}), and \cite{ChPW} are not suitable to model the physical problems studied in \cite{CMKTA} and \cite{ChPW}, respectively.

%point out a curiosity. Reference \cite{AEK} is a frequently cited publication. Based on \cite{AEK}, in \cite{CMKTA} solutions are presented, which are experimentally confirmed in \cite{VSN} ("good agreement between theory and experiment"). 

%It should be also emphasized that ansatz (1b)
%is compatible with the CNLSE (1a) in the nongeneric cases as outlined in  \cite{Se}, Section 4. 

%\begin{figure}[h!]
%\includegraphics[width=22cm]{Fig.1-III}
%\caption{Imaginary part $P(x, t)$ of the CNLSE (1a) in dependence of the variables $x$ and $t$.}
%\end{figure}

%\begin{figure}[h!]
%\includegraphics[width=16cm]{Fig2-III}
%\caption{Phase diagram $\{(Q_x(x,1))^2, Q(x,1)\}$ for admissible $Q_0$ ($Q_0=0$ been selected in Eq.(7)).}
%\end{figure}

%\begin{figure}[h!]
%\includegraphics[width=16cm]{Fig3-III}
%\caption{Imaginary part $P(x,t)$ of the CNLSE (1) with (4) and (7) for $t=1$ according to Eq.(8).}
%\end{figure}

\section*{References}

%Create the reference section using BibTeX:


\begin{thebibliography}{99}
\bibitem{Footnote} There are more than 400 citations of \cite{AEK} in the Web of Science.
\end{thebibliography}
\end{document}